\documentclass[aps,prl,showpacs,groupedaddresss,superscriptaddress, twocolumn]{revtex4}
\usepackage{graphicx}
\usepackage{amsmath}
\usepackage{amsfonts,amssymb}
\usepackage{bm}
\usepackage{color}
\begin{document}

\title{Manipulation and Characterization of the Valley Polarized Topological Kink States in Graphene Based Interferometers}
\author {Shu-guang Cheng}
\affiliation{Department of Physics, Northwest University, Xi'an 710069, China}
\affiliation{Shaanxi Key Laboratory for Theoretical Physics Frontiers, Xi'an 710069, People's Republic of China}
\author {Haiwen Liu}
\affiliation{Center for Advanced Quantum Studies, Department of Physics, Beijing Normal University, Beijing 100875, China}
\author {Hua Jiang}
\email{jianghuaphy@suda.edu.cn}
\affiliation{College of Physics, Optoelectronics and Energy, Soochow University, Suzhou, 215006, China}
\affiliation{Institute for Advanced Study, Soochow University, Suzhou, 215006, China}
\author {Qing-Feng Sun}
\email{sunqf@pku.edu.cn}
\affiliation{International Center for Quantum Materials, School of Physics, Peking University, Beijing 100871, China}
\affiliation{Collaborative Innovation Center of Quantum Matter, Beijing 100871, China}
\affiliation{CAS Center for Excellence in Topological Quantum Computation, University of Chinese Academy of Sciences, Beijing 100190, China}
\author {X. C. Xie}
\affiliation{International Center for Quantum Materials, School of Physics, Peking University, Beijing 100871, China}
\affiliation{Collaborative Innovation Center of Quantum Matter, Beijing 100871, China}
\affiliation{CAS Center for Excellence in Topological Quantum Computation, University of Chinese Academy of Sciences, Beijing 100190, China}

\begin{abstract}
Valley polarized topological kink states, existing broadly in the domain wall of hexagonal lattices systems, are identified in experiments, unfortunately, only very limited physical properties being given. Using an Aharanov-Bohm interferometer composed of domain walls in graphene systems, we study the periodical modulation of pure valley current in a large range by tuning the magnetic field or the Fermi level. For monolayer graphene device, there exists one topological kink state, and the oscillation of transmission coefficients have single period. The $\pi$ Berry phase and the linear dispersion relation of kink states can be extracted from the transmission data. For bilayer graphene device, there are two topological kink states with two oscillation periods. Our proposal provides an experimental feasible route to manipulate and characterize the valley polarized topological kink states in classical wave and electronic graphene-type crystalline systems.
\end{abstract}
\pacs{72.80.Vp, 72.10.-d, 73.20.At}
\maketitle
{\it Introduction--}
Topological kink states broadly exist in the domain walls of magnetic topological insulators\cite{R03,R04}, hexagonal lattice materials, etc.\cite{R10,R16,R18,R13, R14,R11,R17,R19,R20,R21,R22,R08,R09,R07,DWs,DWs2,Parittion4, MoS2,R05,add1,add2,add3,R06,Optic1,Optic2} The recent experimental observations of kink states in bilayer graphene have generate great interests in exploring the exotic properties of such states.\cite{R08,R09,R07,DWs,DWs2,Parittion4} However, the kink states are restricted in a very narrow region, which makes it rather difficult for characterizing with common techniques, such as APRES. By now, the approaches of STM, transport and infrared measurement can only prove the existence of kink state. The pseudospin-momentum locking property, the band structure and even the number of kink states have not been determined in experiments yet. Similar problems also exist in $\rm{MoS_2}$,\cite{MoS2} or the monolayer-graphene-like classical wave systems.\cite{R05,add1,add2,add3,R06,Optic2,Optic1}

The valley polarized states can be used for fabricating valley filters, a key device for valleytronics applications.\cite{R12,R13,R21,R22,R14,R15} The domain walls in graphene systems, which host valley polarized topological kink states, can serve as valley filters.\cite{R21,R22} Nevertheless, in a single domain wall the valley current cannot be easily manipulated. Based on a current splitter composed of two crossed domain walls, the current partition rule of kink state is investigated, indicating the manipulation of valley polarized kink states through a splitter.\cite{Parittion4,R23,R24,Parittion3} The control of current in such a splitter is still difficult since the morphology of domain wall is unchangeable when the devices is fabricated, thus prohibit the manipulation of the kink states.

For the applications of valleytronics, the manipulation of valley polarized current conveniently is essential. Very recently, a domain network in a bilayer graphene is observed experimentally,\cite{DWs2}, which makes the interference of the kink states possible. We adopt this platform for the characterization and manipulation of kink states through quantum interference.

\begin{figure}
\includegraphics[width=\columnwidth,viewport=20 150 370 287,clip]{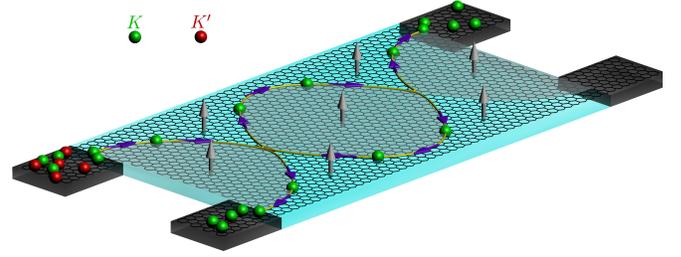}
\caption{The schematics of kink states interferometer under magnetic field. Four terminals are attached to the central region. In the central region, the domain walls (the yellow pathes) are located at the interface of graphene with different inversion symmetry. The purple arrows indicate the propagation direction of the valley $K$ electrons and the grey vertical arrows are for the magnetic field.
}\label{Fig1}
\end{figure}

In this Letter, we study the quantum interference of the Aharanov-Bohm (AB) interferometer\cite{DWs2,piphase} composed of topological kink states locating at the domain walls of graphene systems [see Fig.\ref{Fig1}]. Both the magnetic field and the gate voltage can be utilized to manipulate these kink states, which only allow for the valley polarized propagation and interference. The magnitude of valley polarized current can be adjusted periodically in a wide range by varying the Fermi energy or a magnetic field. The number of kink states can be obtained from the oscillation pattern of the transmission coefficients: for monolayer graphene interferometer, there is one kink state and one oscillation period; while for bilayer graphene system two kink states and two periods exist. Specifically for the former case, a $\pi$ Berry phase and a linear band structure of kink state can be obtained from the transport data. These exotic kink states can broadly exist in graphene-type systems including their classical wave cousins,\cite{R05,add1,add2,add3,R06,Optic1,Optic2} and thus can be realized in present experiments.

\begin{figure}
\includegraphics[width=\columnwidth, viewport=110 62 520 295, clip]{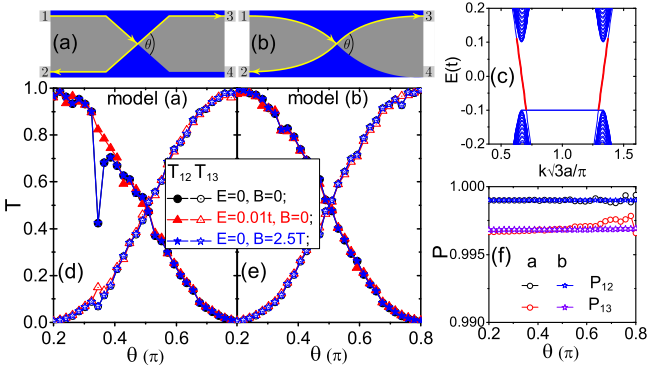}
\caption{The schematics for kink state splitters: straight line (a) and cosine shaped (b). (c) The band structures for kink states in graphene model. (d-e) Angle dependence of $T_{12}$ and $T_{13}$ for model (a) and (b) for different $E$ and $B$, respectively. (f) The angle dependence of valley polarization of $T_{12}$ and $T_{13}$ when $E=0$, $B=0$ for model (a) and (b). The devices' sizes are $L=600a$ (width) and $W=300\sqrt{3}a$ (length).
}\label{Fig2}
\end{figure}

{\it Model and methods--}
Four terminals graphene nanoribbon based devices with a single splitter [see Fig.\ref{Fig2} (a) and (b)] or two splitters [see Fig.\ref{Fig1}] are investigated. Under a small voltage bias between terminal $1$ and other terminals, only electrons of valley $K$ can flow from terminal $1$ into the central region. The current is partitioned once at the splitter in devices of Fig.\ref{Fig2} (a) and (b). In the device of Fig.\ref{Fig1}, the current is partitioned many times and the multi-beam interference happens.

The realization of monolayer graphene based domain walls is difficult in experiment. In recent works, the valley dependent transports are observed in hexagonal lattice of classical wave systems.\cite{R05,add1,add2,add3, R06,Optic2} The physical picture behind is the the same to the well known monolayer graphene which we will focus in the following study. In the tight-binding representation, the model Hamiltonian reads:\cite{R20,R22}
\begin{equation}\label{EQ1}
H_m=\sum_{i} \varepsilon_i c^{\dagger}_ic_i+\sum_{ \langle ij \rangle} t (e^{i\phi_{ij}}c^{\dagger}_i c_j+e^{-i\phi_{ij}}c^{\dagger}_j c_i)
\end{equation}
with $c^\dagger_i$ and $c_i$ are the creation and annihilation operators of electrons at site $i$, respectively. The second term represents the nearest neighbor coupling with energy $t$. The uniform perpendicular magnetic field exists only in the central region and is accounted by the phase factor $\phi_{ij}$. Here $\varepsilon_i$ is the on-site energy at site $i$ to generate spatial inversion asymmetry and hence defines the domain walls.\cite{R10,R22} For monolayer graphene, there are two set of sublattices, namely $A$ and $B$. In the blue (grey) region [see Fig.\ref{Fig2} (a) and (b)], $\varepsilon_A=U(-U)$ and $\varepsilon_B=-U(U)$. In the four terminals, $\epsilon_i$ is shifted by gate voltage $V$ away from the neutral point, which guarantees that there are dozens of states for each valley at the Fermi level.

Adopting the transfer matrix method, the transmission coefficients are calculated.\cite{Ando,R15} Follow the standard procedure, each transmission coefficient $t^{ij}_{mn}$ from state $i$ in terminal $m$ to state $j$ in terminal $n$ is obtained. The valley resolved transmission coefficients are accessed by collection $t^{ij}_{mn}$ in two separate valleys ($K$ and $K'$):
\begin{equation}\label{EQ2}
T^{K_1K_2}_{mn}(E, B)=\sum_{i\in K_1}^m\sum_{j\in K_2}^nt^{ij}_{mn}
\end{equation}
with $E$ the Fermi energy and $B$ the magnetic field. The total transmission coefficient is $T_{mn}=T^{KK}_{mn}+T^{KK'}_{mn}+T^{K'K}_{mn}+T^{K'K'}_{mn}$ and the corresponding valley polarization is $P_{mn}=(T^{KK}_{mn}+T^{KK'}_{mn}-T^{K'K}_{mn}-T^{K'K'}_{mn})/T_{mn}$.
In the numerical calculation we use $U=0.1t$ and $V=0.2t$ with $t$ the energy unit in Eq. \ref{EQ1}. Under the gauge transformation, the magnetic field $B$ is related to $\phi$ by $B={2h}/{3\sqrt{3}a^2\pi e}$ with $a$ the C-C bond length.
At zero temperature, $I_{mn}$ is proportional to $T_{mn}$ due to Landauer-Buttiker formula, so we only concern $T_{mn}$ in the following discussion.

{\it Single splitter--}
The band structure of the kink states for monolayer graphene model is displayed in Fig.\ref{Fig2} (c). The propagation direction of the kink states at different valley is opposite to each other. First the transport of kink states in devices with a single splitter, shown in Fig. \ref{Fig2} (a) and (b), are investigated. Current injected from terminal $1$, can only transmitted into terminal $2$ and $3$. Note that in the domain wall near terminal $2$, the current is of valley $K$ too because the energy band is reversed.\cite{R16} We find $T_{12}+T_{13}\simeq1$ due to the valley conservation and both $T_{12}$ and $T_{13}$ are of high $K$ valley polarized [see Fig.\ref{Fig2} (f)], signalling nearly pure valley current are obtained in the terminal $2$ and $3$. In the following we only discuss the magnitudes of transmission coefficients. The angle dependent partition rule in Ref \cite{R23} is reproduced in Fig. \ref{Fig2} (d) and (e) with nonzero $E$ and $B$. The partition of valley current is only sensitive to the intersection angle at the cross point and is independent of the Fermi level (near the neutral points), the magnetic field, or the specific line shape of the splitter.

\begin{figure}
\begin{center}
\includegraphics[width=\columnwidth, viewport=100 113 665 443, clip]{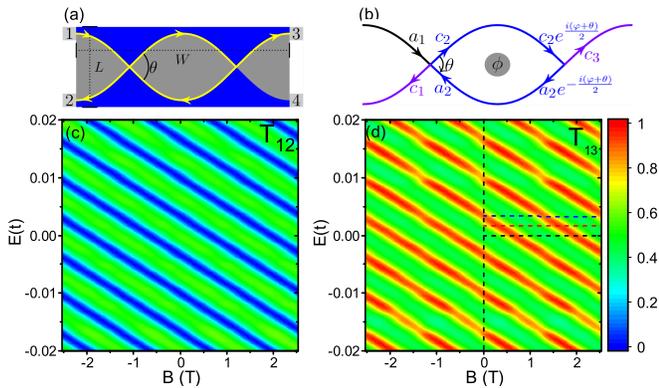}
\caption{The schematics for kink states' interferometer (a) and wave propagation (b). Two dimensional map of $E$ and $B$ dependence of $T_{12}$ (c) and $T_{13}$ (d). The sample sizes are $L=600a$, $W=600\sqrt{3}a$ and $\theta$ is the same to Fig. \ref{Fig2} (b).
}\label{Fig3}
\end{center}
\end{figure}
{\it Monolayer graphene interferometer--}
Now we focus on interferometer formed by domain walls with two splitters as shown in Fig.\ref{Fig3} (a). The incoming wave from terminal $1$, at the left splitter, is partitioned to terminal $2$ and the upper arm of interferometer. At the right splitter, again the wave is split into terminal $3$ and the lower arm. Then, the wave along the lower arm meets the first splitter and is split for the third time. The process happens repeatedly and the currents flow into terminal $2$ and $3$ are the interference of multibeams. The Fermi level $E$ and magnetic field $B$ dependence of $T_{13}(E,B)$ is shown in Fig.\ref{Fig3} (c) and (d). Also $T_{12}+T_{13}\simeq1$ due to the weak backscattering. Significantly, both $T_{12}$ and $T_{13}$ are modulated by $B$ and $E$ periodically in a wide range (e.g. $T_{13}^{max/min}=0.93/0.41$ and $T_{12}^{max/min}=0.57/0$). Note that both $T_{12}$ and $T_{13}$ are nearly valley polarized, so the device is a good controllable valley filter.

\begin{figure}
\includegraphics[width=\columnwidth, viewport=111 232 500 463, clip]{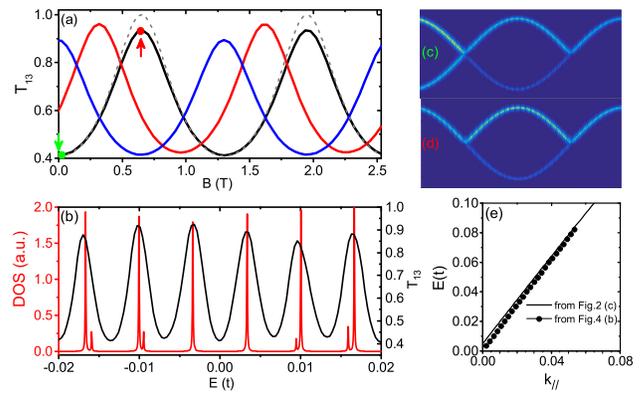}
\caption{$B$ (a) and $E$ (b) dependence of $T_{13}$ along the dashed lines in Fig. \ref{Fig3} (d). The dashed curve in (a) is from Eq.\ref{EQ3}. The red curve in (b) is the DOS of an isolated kink states circle. (c-d) $\rho(\mathbf{r},E)$ for point marked in (a). (e) The energy band of kink state in graphene around valley $K$.
}\label{Fig4}
\end{figure}

The details of $T_{13}$ are displayed in Fig.\ref{Fig4} (a) and (b). In Fig.\ref{Fig4} (a), two main characters are observed: i) $T_{13}$ shows a minimum at $B=0$ and $E=0$; ii) the oscillation of $T_{13}(E_i,B)$ vs. $B$ is shifted periodically as $E_i$ changes. The magnetic flux enclosed by the interferometer [from the period of $T_{13}$] is about $h/e$, indicating the well performance of AB interference. To better understand the characters, the redistribution of valley $K$ electrons from terminal $1$ in the interferometer is pictured by the non-equilibrium local density of states (DOS) [see Fig.\ref{Fig4} (c-d) with the parameters marked in Fig.\ref{Fig4} (a)]. It is calculated by $\rho(\mathbf{r},E)= [\mathbf{G}^r\mathbf{\Gamma}_L\mathbf{G}^a]_{\mathbf{r}}/2\pi$ with $\mathbf{G}^{r/a}$ the Green's function and $\mathbf{\Gamma}_L$ the line-width function.\cite{R20} $\rho(\mathbf{r},E)$ is large around the domain walls and are infinitesimal otherwise, indicating the well formation of interferometer. Moreover, in Fig.\ref{Fig4} (c) ($T^{min}_{13}$), there is current flows into both terminal $2$ and $3$. In Fig.\ref{Fig4} (d), the current flows into terminal $3$ with terminal $2$ blocked. So the interference of valley current is well tuned by the magnetic field.

Fig.\ref{Fig4} (b) shows the $T_{13}$ vs. $E$ relation when $B=0$. $T_{13}$ varies periodically, in analogy to Fig.\ref{Fig4} (a), indicate that the Fermi level plays a similar role as that of magnetic field, i.e. providing an extra phase. Physically the wave function is: $\Psi(\bm{r})=\Psi(r_{\perp})e^{ik_{\parallel}l}$ with $k_{\parallel}$ the momentum. When $E=0$, $k_\parallel$ is zero and the propagation of wave does not contribute extra phase (see Ref. \cite{Supp} Fig. S1). However, when $E\neq0$, $k_\parallel l$ contributes a phase to the wave function and changes the interference. Since the wave in the interferometer can only propagate clockwise, the extra phases of transmission amplitude acquired from the magnetic field and the Fermi level, are similar. To explore the peaks' nature of $T_{13}$, the local DOS of an isolated kink circle [the blue path in Fig.\ref{Fig3} (b)] is shown in Fig.\ref{Fig4} (b). The peaks' positions of $T_{13}$ and DOS are the same. It means the peaks of $T_{13}$ are the result of resonance tunneling through the kink state circle. Interestingly, we find the zero energy is located in the middle of two peaks. It is from the $\pi$ Berry phase of kink state. The wave function of kink state $\Psi(r_{\perp})$ is of two components (pseudospin), which bears the pseudospin-momentum locked topological nature.\cite{Shen} After evolve along a closing circle, acquires a $\pi$ Berry phase.\cite{Supp} Thus the Bohr-Sommerfeld quantization condition $k_\parallel l_0=2n\pi$ is modified to $k_\parallel l_0=2n\pi+\pi$ ($l_0$ the circumference of the interferometer and $k_{\parallel}$ the resonant tunneling wave vector), which means the resonant peak is not located at the zero energy. So the $\pi$ Berry phase can be measured from the features of $T_{13}$. Besides, the dispersion relation of kink states can also be extracted from $T_{13}$. For example, in Fig.\ref{Fig4} (b), there are $n$ peaks from $E=0$ to $E_n$ and the corresponding momentum is $k_{\parallel,n}=(2n+1)\pi/l_0$. In Fig.\ref{Fig4} (e), $E_n\sim k_{\parallel,n}$ relation is shown, in good agreement with the dispersion relation adopted from Fig.\ref{Fig2} (c).

To clarify the above characters, the scattering matrix method is adopted.\cite{MT1,MT2} At the left splitter, the amplitudes of the incident and outgoing waves are indicated by $a_1/a_2$ and $c_1/c_2$, respectively [see Fig.\ref{Fig3} (b)]. Assuming no backscattering, they are related by
$(c_2,c_1)^\mathrm{T}=S(a_1,a_2)^\mathrm{T}$ with $S=\left(
  \begin{array}{cc}
   \sqrt{1-\alpha}e^{-i{\theta}/{2}} & \sqrt{\alpha}e^{i(\pi-\theta)/{2}} \\
 \sqrt{\alpha}e^{i(\pi-\theta)/{2}}  & \sqrt{1-\alpha}e^{-i{\theta}/{2}}\\
  \end{array}
\right)$ the scattering matrix. In $S$, $\sqrt{\alpha}$ and $\sqrt{1-\alpha}$ are the amplitudes of wave from terminal $1$ to $2$ and the upper arm of interferometer, respectively. The Berry phase effect is also accounted, e.g. from $a_1$ to $c_2$, a Berry phase of $e^{-i{\theta}/{2}}$ is acquired when the momentum direction rotates an angle $\theta$ anticlockwise. In the right splitter,
$(c_3, a_2 e^{-i({\varphi+\theta})/2})^\mathrm{T}=
S(c_2e^{i(\varphi+\theta)/2}, 0)^\mathrm{T}$.
Here $\varphi=\phi+k_{\parallel}l_0$ is the combination of the magnetic phase and the dynamic phase. Finally, we have
\begin{eqnarray}\label{EQ3}
\begin{cases}
T_{12}(\alpha,\varphi)=|c_1|^2=\frac{4\alpha\cos^2\frac{\varphi}{2}}{(1-\alpha)^2+4\alpha\cos^2\frac{\varphi}{2}} \\
T_{13}(\alpha,\varphi)=|c_3|^2=\frac{(1-\alpha)^2}{(1-\alpha)^2+4\alpha\cos^2\frac{\varphi}{2}}
\end{cases}
\end{eqnarray}
with $T_{12}+T_{13}=1$ by using $a_1$ as a unit. The result is a Fabry-P\'{e}rot type interference and comes from the multi-beam interference. When $B=0$, $T_{12}$ shows a maximum and $T_{13}$ shows a minimum, the same to Fig.\ref{Fig3}. For the sample size in Fig.\ref{Fig3}, $\theta\simeq0.615\pi$, corresponds to $\alpha\simeq0.22$ [see Fig.\ref{Fig2} (c)]. Substitute this value into Eq. \ref{EQ3} and use $k_{\parallel}=0$, the analytical results is displayed in Fig.\ref{Fig4} (a), in good agreement with the numerical curve for $E=0$. The small discrepancy at $\varphi=\pi$ is due to the back-scattering which is omitted in the analytical approach.

\begin{figure}
\includegraphics[width=\columnwidth, viewport=79 212 612 480, clip]{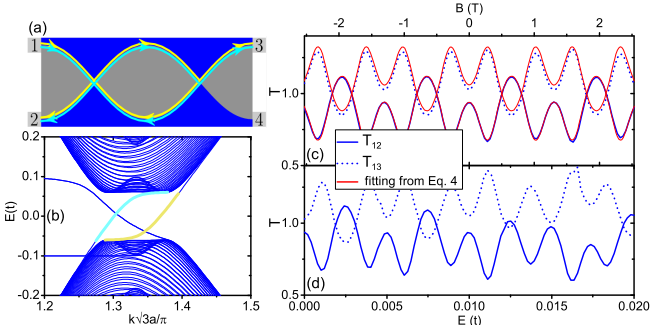}
\caption{(a) The schematics for wave propagation in bilayer graphene model. (b) The energy band of kink states in bilayer graphene around $K$ point. $T_{12}$ and $T_{13}$ vs. $B$ (d) when $E=0$ and vs. $E$ (d) when $B=0$. The red curves are from Eq. \ref{EQ4}.
}\label{Fig5}
\end{figure}

{\it Bilayer graphene interferometer--}
From the technique aspect, kink states interferometers in bilayer graphene are more accessible in experimental.\cite{R08,R09,R07,DWs,DWs2,Parittion4} In the following we investigate the Bernal-stacking bilayer graphene device. The tight binding Hamiltonian is $H_b=H_m^{top}+H_m^{bottom}+\sum_{ij} t_{\perp}(c^\dagger_ic_j+ c^\dagger_jc_i)$. It is constructed by the top and bottom layer of monolayer graphene (Eq.\ref{EQ1}) and the coupling in between. The spatial inversion asymmetry of bilayer graphene is induced by applying an electric field. In our model, it is accounted by the on-site energy difference between two layers. In the blue (grey) region of Fig. \ref{Fig5} (a), $\varepsilon=U (-U)$ in the upper layer and $\varepsilon=-U (U)$ in the bottom layer with $U=0.1t$.
The nearest coupling between $A$ and $B$ atoms in two layers is $t_{\perp}=0.15t$.

Fig.\ref{Fig5} plots the results for bilayer graphene model. In Fig.\ref{Fig5} (c) and (d), both $T_{12}$ and $T_{13}$ vs. $B$ show periodical oscillations and the period is the same to the case for monolayer model because of the same sample size.
Especially, $T_{12}/T_{13}$, proportional to the pure valley current in terminal $2/3$, can be tuned for 1.68/1.53 times ($\frac{T^{max}_{13}}{T^{min}_{13}}$/$\frac{T^{max}_{12}}{T^{min}_{12}}$). So the interference in bilayer graphene has promising application in valley current modulation as well. Different from monolayer graphene model which has only one period, $T_{13}$ in Fig.\ref{Fig5} (c) has two periods. It is ascribed to the double kink states in bilayer model [see Fig.\ref{Fig5} (b)]. When $E=0$, the phases of two kink states are ¦Õ$\varphi_{\pm}=\phi\pm k^0_{\parallel}l$, including a same magnetic phase $\phi$ and non-zero dynamic phases with $\pm k^0_{\parallel}$ the momentum of two kink states (symmetric with respect to $K$ point). So $T_{12}$ and $T_{13}$ are the summation of two separate kink states' transmission. It can be directly obtained by using the scattering method:
\begin{eqnarray}
\begin{cases}
T_{12}(\alpha,\phi)=\frac{4\alpha\cos^2\frac{\varphi_+}{2}}{(1-\alpha)^2+4\alpha\cos^2\frac{\varphi_+}{2}} +\frac{4\alpha\cos^2\frac{\varphi_-}{2}}{(1-\alpha)^2+4\alpha\cos^2\frac{\varphi_-}{2}} \\
T_{13}(\alpha,\phi)=\frac{(1-\alpha)^2}{(1-\alpha)^2+4\alpha\cos^2\frac{\varphi_+}{2}}+ \frac{(1-\alpha)^2}{(1-\alpha)^2+4\alpha\cos^2\frac{\varphi_-}{2}}.
\end{cases}\label{EQ4}
\end{eqnarray}
In Fig.\ref{Fig5} (c), the fitting curves from Eq. \ref{EQ4} are plotted ($\alpha=0.28$ and $k^0_{\parallel} l=1.75$), in close agreement with the numerical results. From Eq. \ref{EQ4}, two periods associate with $\varphi_{\pm}$ are clearly seen. So far the double kink states in bilayer graphene have not yet be verified, e.g. the conductance is smaller than $4e^2/h$.\cite{R08,R09} Thus our proposal can provide a direct evidence to explore it. Besides, the two periods' oscillation exists in the presence of moderate disorder (see Ref.\cite{Supp} Fig.S5). So the double kink states can be confirmed in experiment more easily with our proposal, i.e. no quantized conductance required. Fig.\ref{Fig5} (d) shows the $E$ dependence of $T_{12}$ and $T_{13}$ when $B=0$. The two period feature still hold. But there exist irregular oscillation pattern which should be ascribed to the non-linear dispersion of kink states in bilayer graphene [see Fig.\ref{Fig5} (b)]. Fig.\ref{Fig5} (d) also demonstrate that valley current can be tuned by the Fermi level in bilayer graphene. From Eq.\ref{EQ3} and Eq.\ref{EQ4}, the modulation of valley current is dominated by the interference phase, so above discussions hold true for interferometers of
irregular shapes.\cite{DWs2}

{\it Conclusion--}
In conclusion, an AB interferometer is proposed to characterize and manipulate the topological valley kink states. The output of the current is perfectly valley polarized and changes periodically in a large range when sweeping the Fermi energy or the magnetic field. It provides a versatile and high efficiency way to manipulate valley degrees of freedom. For monolayer graphene, due to a single kink state, there is only one period in the transmission coefficient. The $\pi$ Berry phase and the linear band of kink state can be extracted from the transport measurements. However, for the bilayer graphene, there are two kink states. The transmission coefficients modulated by a magnetic flux shows two periods. Our proposal can also be used in experiments to characterize the topological nature of the kink states.

We thank the stimulus discussion with L. He and Z. W. Shi. We also acknowledge the discussion with Z. H. Hang from the experimental aspect and the related work is under processing. This work was supported by NSFC under Grants Nos. 11674264, 11534001, 11574007 and 11674028, NSF of Jiangsu Province, China (Grant No. BK20160007), National Key R and D Program of China (2017YFA0303301), NBRP of China (2015CB921102) and the Key Research Program of the Chinese Academy of Sciences (Grant No. XDPB08-4).

\end{document}